\documentclass{iaus}

\usepackage{graphicx}

\title[AO observations of Exoplanets, Brown Dwarfs, and Binary Stars] 
{Adaptive Optics Observations of Exoplanets, Brown Dwarfs, \& Binary Stars}

\author[Sasha Hinkley]  {Sasha Hinkley$^1$ }

\affiliation{$^1$Sagan Fellow, California Institute of Technology, Mail Code 249-17, 1200 E. California blvd., Pasadena, CA 91125 \\ email: {\tt shinkley@astro.caltech.edu} }

\pubyear{2011}
\volume{282} 
\pagerange{???}
\jname{From Interacting Binaries to Exoplanets: Essential Modeling Tools}
\editors{A.C. Editor, B.D. Editor \& C.E. Editor, eds.}

\begin{document}

\maketitle

\begin{abstract}
The current direct observations of brown dwarfs and exoplanets have been obtained using instruments not specifically designed for overcoming the large contrast ratio between the host star and any wide-separation faint companions.  However, we are about to witness the birth of several new dedicated observing platforms specifically geared towards high contrast imaging of these objects.  The Gemini Planet Imager, VLT-SPHERE, Subaru HiCIAO, and Project 1640 at the Palomar 5m telescope will return images of numerous exoplanets and brown dwarfs over hundreds of observing nights in the next five years.  Along with diffraction-limited coronagraphs and high-order adaptive optics, these instruments also will return spectral and polarimetric information on any discovered targets, giving clues to their atmospheric compositions and characteristics. Such spectral characterization will be key to forming a detailed theory of comparative exoplanetary science which will be widely applicable to both exoplanets and brown dwarfs.  
Further, the prevalence of aperture masking interferometry in the field of high contrast imaging is also allowing observers to sense massive, young planets at solar system scales ($\sim$3-30 AU)--- 
separations out of reach to conventional direct imaging techniques. Such observations can provide snapshots at the earliest phases of planet formation---information essential for constraining formation mechanisms as well as evolutionary models of planetary mass companions.  As a demonstration of the power of this technique, I briefly review recent aperture masking observations of the HR 8799 system.  Moreover, all of the aforementioned techniques are already extremely adept at detecting low-mass stellar companions to their target stars, and I present some recent highlights. 

\end{abstract}

\section{Introduction}
In recent years, astronomers have identified more than 400 planets outside our solar system, launching the new and thriving field of exoplanetary science (Marcy \etal~2005).  The vast majority of these objects have been discovered indirectly by observing the variations induced in their host star's light.  The radial velocity surveys can  provide orbital eccentricity, semi-major axes, and lower limits on the masses of companion planets while observations of transiting planets can provide fundamental data on planet radii and limited spectroscopy of the planets themselves.  However, studying those objects out of reach to the radial velocity and doppler methods will reveal completely new aspects of exoplanetary science in great detail.    {\it Direct imaging} of exoplanets provides a complementary set of parameters such as photometry (and hence luminosity), as well as detailed spectroscopic information.

\begin{figure}
\centering
\resizebox{13.5cm}{!}{\includegraphics[angle=-90]{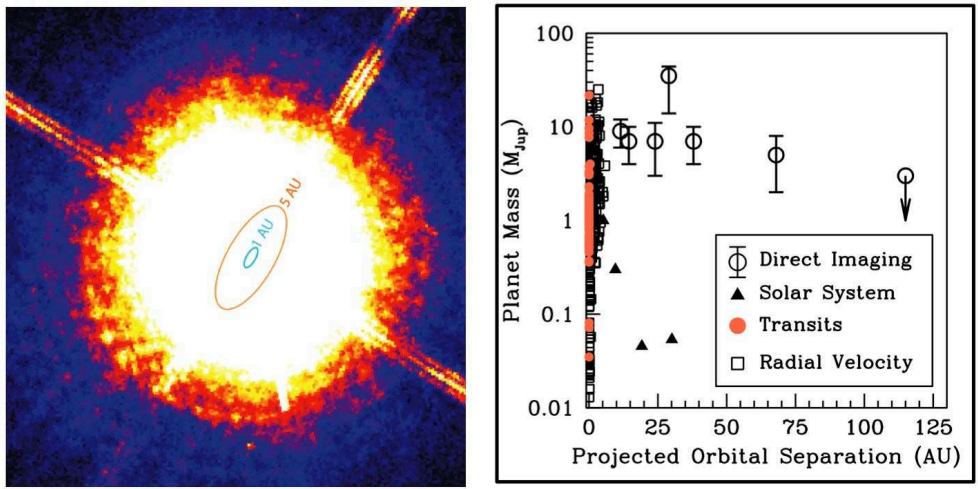}}
\caption[]{{\it Left:} A schematic demonstration taken from \cite{oh09} demonstrating the challenges associated with high contrast imaging. Hypothetical orbits of a Jupiter and an Earth have been overlaid on the Point Spread Function of a nearby star. Planetary mass companions typically have angular separations of a fraction of an arcsecond, and contrast ratios spanning several orders of magnitude ($\sim$10$^4$-10$^7$).  {\it Right:} The Mass vs. Orbital separation for planetary mass companions detected through transits (filled circles), radial velocity measurements (squares), solar system planets (triangles), as well as those objects detected through direct imaging (open circles). }
\label{highcontrast}
\end{figure}

Moreover, the direct imaging of planetary mass companions will allow researchers to more fully probe the mass-separation parameter space (Figure~\ref{highcontrast}, right panel) occupied by these objects (Oppenheimer \& Hinkley 2009). At young ages (\S 3), the orbital placement of planetary mass companions serves as a birth snapshot, lending support to models (Pollack \etal~1996) that may be more efficient at building a massive core at only a few AU, or models that allow for the formation of massive objects at tens of AU though fragmentation of the gaseous disk (Boss 1997).  Finally, imaging of multi-planet systems at any age will serve as dynamical laboratories for studying the planetary architectures. 

\begin{figure}
\centering
\resizebox{14.5cm}{!}{\includegraphics[angle=90]{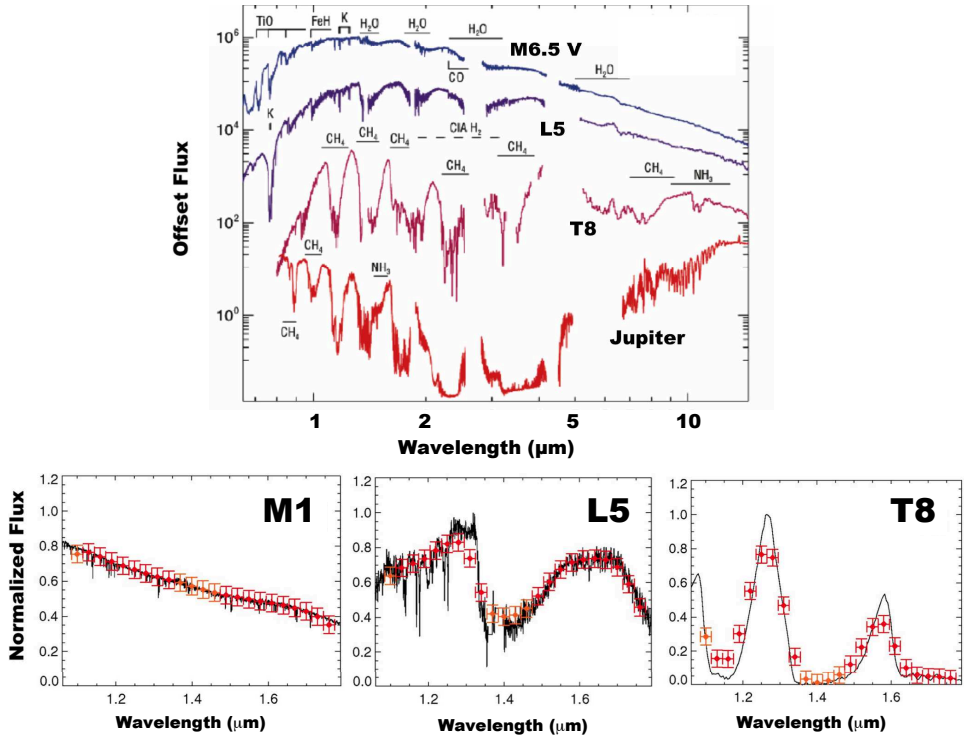}}
\caption[]{{\it Top Panel:} Example Spectra for M, L, and T-dwarfs, as well as Jupiter with prominent absorption bands marked, taken from Marley \etal~(2009).  {\it Lower Panels:}  The points with error bars are simulated measurements gathered with a low resolution ($\lambda/\Delta\lambda$$\sim$45) spectrograph, shown along with higher resolution spectra from NASA IRTF (black curves) for the labelled spectral types. Such low spectral resolution measurements will be obtained by GPI, SPHERE and Project 1640, and can easily sample broad absorption features present in late type stars and exoplanets.  Lower panels are taken from Rice \etal~(2011) in prep. Images used courtesy of Michael Cushing, Mark Marley, and Emily Rice. }
\label{spectra}
\end{figure}

Perhaps just as essential, spectroscopy provides clues to the atmospheric chemistry, internal physics, and perhaps may even shed light on non-equilibrium chemistry associated with these objects.  More robust classification schemes for planets in general will arise from observing as many planets as possible at different ages, in different environments, and with a broad range of parent stars. Figure~\ref{spectra} provides an illustration of the diversity of the spectra of late-M, L and T dwarfs as well as a spectrum of Jupiter showing broad absorption bands due to e.g. H$_2$O, CH$_4$, and NH$_3$.  Spectral characterization of such objects can be accomplished even with the somewhat low spectral resolution ($\lambda/\Delta\lambda\sim$30-50) of the instruments described in this review. These pieces of information not only reveal the detailed properties of the objects themselves, they serve as key benchmarks for competing evolutionary models describing the thermal and atmospheric properties of these objects.


It is often overlooked, but should be emphasized, that the recent spectacular images (Marois \etal~2008, Kalas \etal~2008, and Lagrange \etal~2010)  of Jovian planets were obtained with Adaptive Optics, hereafter ``AO'', systems and infrared cameras not specifically designed for the task of overcoming the high contrast ratio between the planets and their host stars. These were obtained using existing instrumentation, but with observing and data reduction strategies customized for high contrast imaging. Further, the recent L-band images of HR 8799 and $\beta$ Pic b (Marois \etal~2010, Lagrange \etal~2010) were obtained without the use of a  coronagraph! These studies have demonstrated that direct imaging of planetary mass companions as well as disks (e.g. Oppenheimer \etal~2008, Hinkley \etal~2009) is now a mature technique and may become routine using ground-based observatories. More so, the handful of coming instruments dedicated to detailed spectroscopic characterization of planetary mass companions may make these kinds of discoveries routine, initiating an era of comparative exoplanetary science. 

\subsection{The Challenge of High Contrast Imaging}

The major obstacle to the direct detection of planetary companions to nearby stars is the overwhelming brightness of the host star.  If our solar system were viewed from 20 pc, Jupiter would appear $10^8-10^{10}$  times fainter than our Sun in the near-IR (Barraffe \etal~2003) at a separation of 0.25$^{\prime\prime}$, completely lost in its glare (See Figure~\ref{highcontrast}, left panel). The key requirement is the suppression of the star's overwhelming brightness through precise starlight control (Oppenheimer \& Hinkley 2009). 

	A promising method for direct imaging of stellar companions involves two techniques working in conjunction.  The first, high-order AO, provides control and manipulation of the image by correcting the aberrations in the incoming stellar wave front caused by the Earth's atmosphere.  AO has the effect of creating a nearly diffraction-limited point spread function, with the majority of the stellar flux concentrated in this core.  Second, a Lyot coronagraph (Sivaramakrishnan \etal~2001) suppresses this corrected light.  Together, these two techniques can obtain contrast levels of $10^4$-$10^5$ at 1$^{\prime\prime}$ (Leconte \etal~2010).  Improvements in coronagraphy, specifically the apodization of the telescope pupil (Soummer 2005), as well as post-processing to suppress speckle noise (Hinkley \etal~2007, Crepp et al 2011), can significantly improve the achieved contrast, especially at high Strehl ratios.  Below we briefly describe some of the instrumentation being built at the time of this writing, or currently in place on large telescopes.  

\section{New High Contrast Instrumentation for Spectroscopic Studies of Exoplanets \& Brown Dwarfs}

\subsection{Southern Hemisphere: The Gemini Planet Imager, SPHERE, and NICI}

\begin{figure}
\centerline{
\scalebox{0.5}{%
\includegraphics[angle=90]{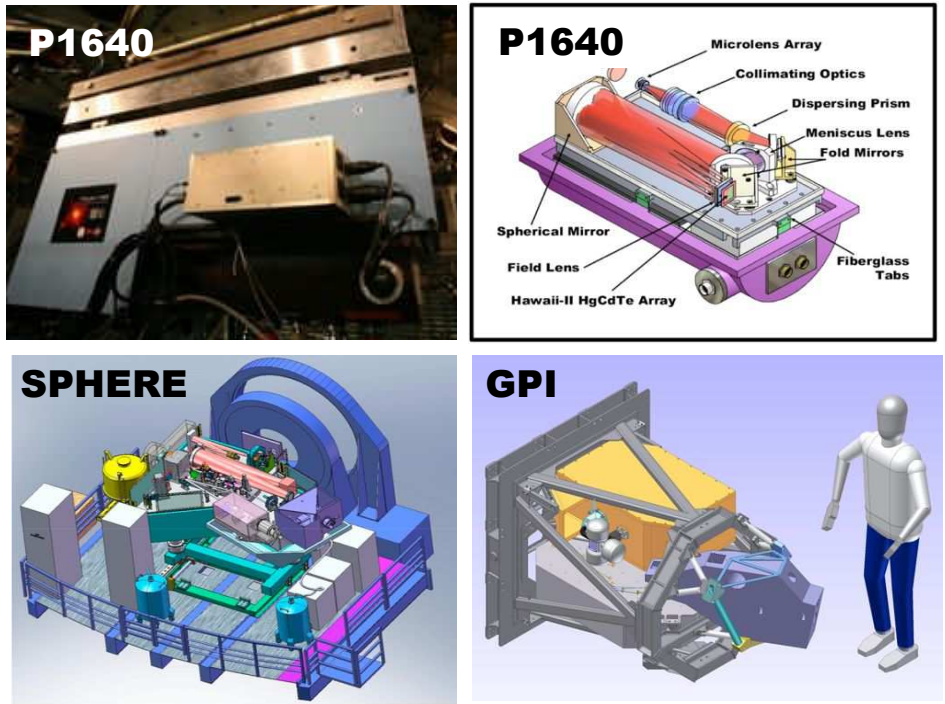}
}
}
\caption[]{{\it Top left:} The Project 1640 high contrast imager as described in Hinkley et al (2011c), and installed at the 200-in Hale Telescope.  {\it Top right:} the internal optical layout of the Project 1640 Integral Field Spectrograph. The SPHERE and GPI designs are shown on the lower left and right, respectively. Each of these instruments are designed to image exoplanetary mass companions at contrasts of $10^7$ or greater, as well as return spectroscopic information.}
\label{projects}
\end{figure}

In Figure~\ref{projects} we show design drawings for two dedicated high contrast instruments with the goal of imaging and studying in detail extrasolar planets:  the Gemini Planet Imager (Macintosh \etal~2008), hereafter ``GPI'', and the Spectro-Polarimetric High-contrast Exoplanet Research (Beuzit \etal~2008), hereafter ``SPHERE.'' 

{\underline{\it The Gemini Planet Imager}}: GPI will be deployed to the 8m Gemini South telescope in 2012 to begin a survey of several hundred nearby stars in young associations with ages $\sim$10-100 Myr. With contrast goals of $10^7$-$10^8$, the instrument will gather spectroscopic and polarimetric information on the detected exoplanets. The core instrument-package is: a 1500-subaperture MEMS AO system; an apodized-pupil Lyot coronagraph; a post-coronagraph wave front calibration system; and a low resolution ($\lambda/\Delta\lambda$$\sim$10-100) integral field spectrograph covering the $J$, $H$, and $K$-bands with polarimetric capabilities. 

{\underline{\it SPHERE}}: SPHERE at VLT (Beuzit \etal~2008) will be a dedicated high contrast imaging instrument with similar science goals as GPI.  With very small inner working angles ($\sim$100 mas), SPHERE will survey nearby young associations (10-100 Myr) within 100 pc. In addition, the survey will target young active F/K dwarfs and the nearest stars within 20 pc to sense reflected light directly from the planets. The project also intends to monitor stars showing long-term radial velocity trends, astrometric planet candidates, as well as known multi-planet systems.  
%

{\underline{\it NICI}}: The Near Infrared Coronagraphic Imager, ``NICI'' (Liu \etal~2010),  at the 8m Gemini-South telescope is using a variety of high contrast techniques to monitor $\sim$300 young and nearby stars.   NICI utilitzes an 85-element curvature wave front system, Lyot coronagraphy, Simultaneous Differential Imaging (``SDI'') around the methane band, and Angular Differential Imaging (``ADI'') for a variety of observing options. NICI will be an effective forerunner for the next generation of high contrast imaging platforms in the Southern Hemisphere (GPI, SPHERE) with spectroscopic capabiliites.

\subsection{Northern Hemisphere: Project 1640 at Palomar \& Subaru HiCIAO}

{\underline{\it Project 1640 at Palomar:}}  We briefly describe a new instrument which forms the core of a long-term high contrast imaging program at the 200-in Hale Telescope at Palomar Observatory.  The detailed descriptions of the instrumentation can be found in Hinkley \etal~(2008), and Hinkley \etal~(2011b), and we show an image of the instrument mounted on the 200-in telescope in Figure~\ref{projects}.  The primary scientific thrust is to obtain images and low-resolution spectroscopy ($\lambda/\Delta\lambda\sim45$) of brown dwarfs and young exoplanets of several Jupiter masses in the vicinity of stars within 50 pc of the Sun. 

The instrument is comprised of a microlens-based integral field spectrograph (hereafter ``IFS''), an  apodized-pupil Lyot coronagraph, and a post-coronagraph internal wave front calibration interferometer mounted behind the Palomar adaptive optics system. The spectrograph obtains imaging in 23 channels across the $J$ and $H$ bands (1.06 - 1.78 $\mu$m).  
The Palomar AO system is undergoing an upgrade to a much higher-order AO system (``PALM-3000"): a 3388-actuator tweeter deformable mirror working together with the pre-existing 241-actuator mirror.  This system will allow correction with subapertures as small as 8.1cm at the telescope pupil using natural guide stars.  
As part of the project's first phase, the AO system, coronagraph, and IFS achieved contrast of $2\times10^{-5}$ at $1^{\prime\prime}$ (Crepp \etal~2011, Hinkley \etal~2011b). 
We anticipate this instrument will make a lasting contribution to high contrast imaging in the Northern Hemisphere for years. 

Among the early science results, we briefly highlight some intriguing findings regarding massive stellar companions to A-stars.  The mid-A star $\zeta$ Virginis, has been discovered to host a mid-M dwarf companion.  As we discuss in \cite{hob10}, this newly discovered object may explain the anomalous X-ray emission from this A-star.  A comparable scenario involves the star Alcor, a bright mid-A dwarf. A mid-M dwarf companion was found to orbit this star, and was identified through the novel technique of common parallax measurements (Zimmerman \etal~2010).  Lastly, Hinkley \etal~(2011a) analyze the $\alpha$ Oph system: a rapidly rotating A-star with a 0.77 M$_{\odot}$ companion. Monitoring this system over several years allowed us to fit an orbit to the companion astrometry, thereby placing an important dynamical constraint on the mass of the rapidly rotating primary.  

{\underline{\it The Subaru SEEDS Survey:}} In addition to Project 1640 at Palomar, the SEEDS survey at Subaru will utilize instrumentation dedicated to observations of planetary mass companions. Sitting at a Nasmyth focus, and integrated with the HiCIAO camera, the AO system consists of a 188-actuator curvature system, plus a MEMS-based wave front control system, with control and correction of the focal plane wave front in the near-infrared. The techniques of Phase-Induced Amplitude Apodization and aperture masking interferometry will allow access to inner working angles of 20-40 mas (Martinache \& Guyon 2009). The platform is intended to be highly flexible, and will allow the instrumentation to adapt as techniques and understanding advance in the field.

\section{Observing the Youngest Systems with High Angular Resolution}
\subsection{Background}

Observing the primordial population of planets in very young stellar associations or moving groups with ages of a few Myr will provide data on these objects in their pristine state.  Specifically, detailed information about the orbital distribution of these giant planets at the moment of formation is largely missing from many of the discussions of planetary formation mechanisms. The so-called core-accretion model (e.g. Pollack \etal~1996) dictates that the formation timescale at several tens of AUs may become prohibitively long. Under certain conditions, the so-called gravitational instability model, however, may be able to form objects more easily at larger orbital separations.  However, as several researchers have suggested, dynamical processes as discussed e.g. by Veras \etal~(2009), could be responsible for the placement of wide separation companions. Observing these nascent planetary systems as early as possible then will serve as a ``snapshot'' of the true formation landscape and greatly eliminate any confusion about the system's initial conditions caused by dynamical processes. 

Moreover, observing this primordial population, just as the gas disk is dissipating, is a key measure that will lend support to divergent evolutionary models of the planets themselves. Two of the leading evolutionary models, the so-called ``hot-start'' models of \cite{bcb03} and the more recent ``cold-start'' models as discussed by \cite{fms08} are largely unconstrained at very young ages---the evolutionary window where they are most divergent.  Observations of young planetary systems will serve as key benchmark measurements against which these models can be compared.  
However, the youngest stellar systems reside in star forming associations ($\sim$140pc), where a Jupiter-like orbital separation of $\sim$6 AU subtends an angle of 40-50 mas---about the size of the diffraction limit for a 10m telescope. {\it Hence, probing within the diffraction limit of the telescope is absolutely critical.}  

\subsection{Aperture Masking Interferometry}
The technique of aperture masking interferometry (e.g. Tuthill \etal~2000, Ireland \& Kraus 2008), achieves contrast ratios of $\sim$$10^2$ - $10^3$ at very small inner working angles, usually within $\sim$$\frac{1}{3}\lambda/D - 4\lambda/D$ ($\sim$ 20-300 mas for Keck $L'$-band imaging).  Other high contrast techniques, such as polarimetry (e.g. as discussed in Graham \etal~2007, Oppenheimer  \etal~2008, \& Hinkley \etal~2009), can achieve impressively small inner working angles in theory comparable to aperture masking.  Applications of aperture masking, (e.g. Tuthill \etal~2000, Ireland \& Kraus 2008, Hinkley \etal~2011c), employ AO with the use of an opaque mask containing several holes, constructed such that the baseline between any two holes samples a unique spatial frequency in the pupil plane. Moreover, a coronagraphic mask is not used in this observing mode, since it would obscure much of the accessible parameter space, as well as providing a great deal of uncertainty in the astrometry of the host star as discussed e.g. in Digby \etal~(2006). 

This technique is particularly well suited for studies of stellar multiplicity and the brown dwarf desert (e.g. Kraus \etal~2008, Ireland \& Kraus 2008), or detecting giant planets orbiting young stars ($\lesssim$~50-100 Myr).  As an example of the power of this technique, we describe the recent results of \cite{hci11}, targetting the HR 8799 system.  We are able to place upper limits on additional massive companions to the HR 8799 system that would otherwise have been veiled by the telescope diffraction or the associated uncorrected quasi-static speckle noise (e.g. Hinkley \etal~2007) using more conventional direct imaging techniques.

\begin{figure}
\centering
\resizebox{14 cm}{!}{\includegraphics[angle=90]{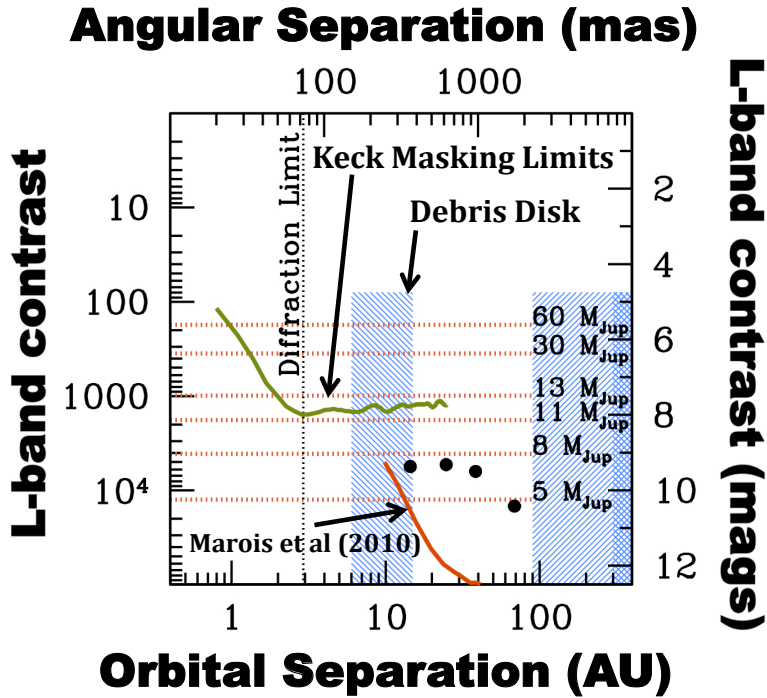}}
\caption[]{{\it The inner 10 AU of HR 8799:} The figure shows detection limits for the HR 8799 system taken from Hinkley \etal~(2011c).  The thick curves show the minimum brightness required for a 99\% confidence detection in our HR 8799 aperture masking data, labelled ``Keck Masking Limits''. Also shown are the \cite{mzk10} sensitivities (C. Marois and B. Macintosh, private communication 2011). 
We also indicate the theoretical diffraction limit at $L'$-band on a 10m telescope (black dotted line), as well as the $L'$-band brightnesses of the four known companions to HR 8799 (black points). The hatched regions represent  the debris disk structures as defined by \cite{srs09} }
\label{nrmcontrasts}
\end{figure}

\subsection{The inner 10 AU of HR 8799}
HR 8799, a $\sim$30 Myr A5V star hosting several planets, presents a challenge for formation models of massive exoplanets.  The recent identification of four $\sim$5-7 M$_{\rm Jup}$ exoplanets (Marois \etal~2008, 2010) reinforces that this system is characterized by a complicated and intriguing architecture. 
Gaining a more complete understanding of additional companions in the HR 8799 system will be essential for understanding the dynamical stability of the system, as well as performing an overall mass census.  

Although we detect no other candidate companions interior to 14 AU, the location of HR 8799e, in Figure~\ref{nrmcontrasts} we show our ability to place constraints on companions more massive than $\sim$11 M$_{\rm Jup}$ within this region.   Details on the calculation of the detection limits can be found in \cite{hci11}.  We achieve our best contrast of 8 mags at the $L'$-band diffraction limit of the Keck telescopes, corresponding to a 3 AU projected orbital separation for HR 8799. This orbital separation is comparable to the ice line boundary for a 30 Myr mid-A star, and these limits place a stringent constraint on any massive inner perturbers that might be responsible for the placement of the four planetary mass companions.  

In addition, the study of the HR 8799 debris disk carried out by \cite{srs09}, indicates the presence of an inner warm ($\sim$150 K) debris belt (see Figure~\ref{nrmcontrasts}).  The outer boundary of this disk is most likely sculpted by the ``e'' component at 14 AU. \cite{srs09} state that few, if any, dust grains exist interior to 6 AU, and postulate that additional planets may be responsible for the clearing.  
Hence, any planetary mass companions responsible for sculpting this inner edge must be $\lesssim$11 M$_{\rm Jup}$

\begin{figure}
\centerline{
\scalebox{0.5}{%
\includegraphics[angle=90]{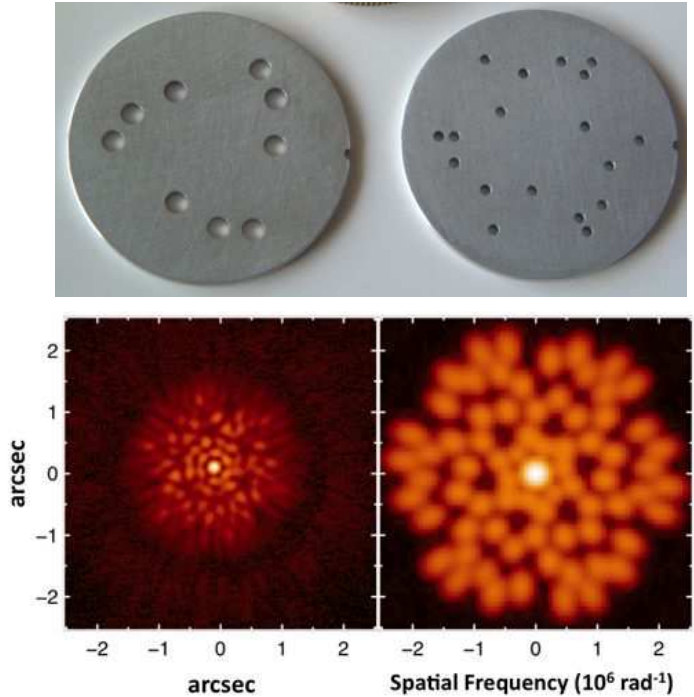}
}
}
\caption[]{{\it Top:} Two of the masks used in the Keck NIRC2 infrared camera for aperture masking interferometry. {\it Lower Left:} A Keck $L'$-band interferogram of the star HR 8799 produced by the nine-hole mask.  {\it Lower Right:} The power spectrum corresponding to the image at lower left. 
}
  \label{data  }
\label{nrmmasks}
\end{figure}

\subsection{The future: multi-wavelength studies}
The science return from this technique will be multiplied when the imaging science camera has multi-wavelength capabilities such as an IFS. This has already been achieved with the Project 1640 Integral IFS and  at Palomar Observatory (Hinkley \etal~2011b, Zimmerman \etal~ 2011, in prep), and will be integrated with the Gemini Planet Imager coronagraph (Macintosh \etal~2008). 
In the more distant future, multi-wavelength aperture masking interferometry will play a prominent role for the planet-finding efforts of JWST, significantly improving on the sensitivity presented here (Sivaramakrishnan \etal~2010, Doyon \etal~2010). 




\end{document}